# Stochastic Computing with Integrated Optics

Hassnaa El-Derhalli
Dept. of Electrical and Computer Engineering
Concordia University
Montreal, Quebec, Canada
h_elderh@ece.concordia.ca

Sébastien Le Beux
Lyon Institute of Nanotechnology
Ecole Centrale de Lyon
Lyon, France
sebastien.le-beux@ec-lyon.fr

Sofiène Tahar
Dept. of Electrical and Computer Engineering
Concordia University
Montreal, Quebec, Canada
tahar@ece.concordia.ca

*Abstract*— Stochastic computing (SC) allows reducing hardware complexity and improving energy efficiency of error resilient applications. However, a main limitation of the computing paradigm is the low throughput induced by the intrinsic serial computing of bit-streams. In this paper, we address the implementation of SC in the optical domain, with the aim to improve the computation speed. We implement a generic optical architecture allowing the execution of polynomial functions. We propose design methods to explore the design space in order to optimize key metrics such as circuit robustness and power consumption. We show that a circuit implementing a 2$^{nd}$ order polynomial degree function and operating at 1Ghz leads to 20.1pJ laser consumption per computed bit.

## I. INTRODUCTION

Stochastic computing (SC) is an approximate computation technique in which numbers are represented as probabilities using random bit-streams [1]. It allows reducing hardware complexity and improving energy efficiency compared to conventional computing circuits. The probabilistic representation of data used in SC are suitable for error tolerant applications such as image and signal processing. SC is also commonly used in application domains where soft errors and process variations are of major concern [2]. The main limitation for the deployment of SC to other domains is the low throughput induced by the intrinsic serial computing of bit-streams. Numerous parallel design techniques have been investigated to overcome the slow computation speed induced by electronic technologies [3]. However, such approach may lead to significant area and power overhead, which thus drastically limits the interest in the SC paradigm.

Nanophotonics is a promising technology to overcome computation throughput limitation thanks to key features of light propagation, namely: low latency and high bandwidth [4]. Integrated optics has been proven efficient to implement microwave filters processors [5] and it has been investigated for the implementation of logical and arithmetic circuits [6]. Hence, the technology could ideally replace CMOS for the execution of SC since it would provide high-speed transmission of serial bit-streams.

However, the implementation of SC circuits using integrated optics faces numerous challenges related to the interactions between optical signals. Indeed, SC implies the use of logic gates in the data path and relies on switching operations to control the signals propagation. A direct transposition of these requirements in the optical domain would lead to the use of active photonic devices controlled using electro-optics effect. However, this solution involves slow electro-optical (E/O) and opto-electrical (O/E) conversions in the data path, which drastically limits the interest of using optical technologies. Hence, a disruptive optical design relying on, non-conventional, all-optical switching effect is mandatory for the execution of SC in the optical domain.

In this paper, for the first time, we propose an optical circuit allowing the execution of SC in the optical domain. Our aim is to combine key advantages of both techniques, namely error resilient computing and high throughput signal transmission. The contributions are the following: We implement a generic optical architecture allowing the execution of polynomial functions. We define an analytical model allowing estimating the optical signal transmission through the circuit. We establish design methods to optimize circuit robustness and laser power consumption according to optical devices characteristics. We conduct a comprehensive study to evaluate the architecture energy efficiency, transmission error resilience and scalability.

The rest of the paper is organized as follows: Section II presents a brief overview of SC techniques and existing optical computing architectures. The proposed architecture is presented in Section III. In Section IV, the implementation and model of the architecture are presented. The experimental results are provided in Section V. Finally, we conclude the paper in Section VI and present some future work directions.

## II. BACKGROUND AND RELATED WORK

In this section, a brief introduction about SC techniques and photonics-based computing architectures is presented.

### A. Stochastic Computing

In SC, binary numbers are converted to random bit-streams that can be interpreted as probabilities. The total number of ones in the bit-stream determines the probability of the binary number. This allows the use of basic logic elements for the hardware design as in [7] and [8], where elementary arithmetic operations such as addition, multiplication, and division were implemented. In [9], a Reconfigurable Stochastic Computing (ReSC) unit was implemented using combinational circuit by converting an arbitrary continuous function to Bernstein polynomial function, which is defined as

$$B(x) = \sum_{i=0}^{n} b_i B_{i,n}(x) \qquad (1)$$

where *x* is the input, *n* is the polynomial degree, $B_{i,n}(x)$ is a Bernstein basis polynomial of degree *n*, and $b_i$ is the Bernstein polynomial coefficient.

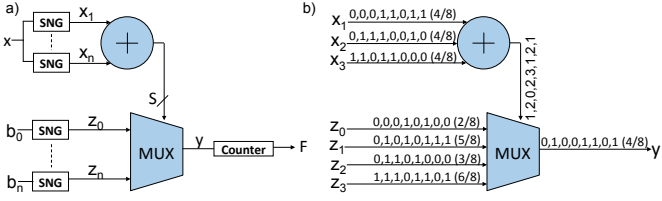

**Fig. 1:** ReSC architecture unit proposed in [9].

As illustrated in Fig. 1(a) the computation is carried out as follows: i) n stochastic number generators (SNG) generate n stochastic bit-stream of data input x from $x_1$ to $x_n$; ii) n+1 SNGs generate bit-streams for the Bernstein polynomial coefficients $z_0$ to $z_n$. iii) the streams of the coefficients are multiplexed to the output according to the sum of input data ($x_1$ to $x_n$) and iv) the number of the received ones are counted to derandomize the data. Fig. 1(b) illustrates the implementation of the Bernstein polynomial $f_1(x) = \frac{2}{8}B_{0,3}(x) + \frac{5}{8}B_{1,3}(x) + \frac{3}{8}B_{2,3}(x) + \frac{6}{8}B_{3,3}(x)$ which is derived from $f_1(x) = \frac{1}{4} + \frac{9}{8}x - \frac{15}{8}x^2 + \frac{5}{4}x^3$, where data x=0.5.

Since the ReSC unit is suitable for numerous error tolerant applications (e.g., image processing, signal processing and neural computation applications), we investigate its transposition in the optical domain.

*B. Silicon Photonics*

The design of optical computing circuits involves optical devices such as Mach-Zehnder Interferometer (*MZI*) and Micro-Ring Resonator (*MRR*), as detailed in the following.

*MZI:* Fig. 2(a) illustrates a 1x1 MZI modulator. The input signal power is equally split and transmitted to two parallel waveguides. On one arm, the signal continues propagating at the speed related to silicon refractive index. On the other arm, the silicon refractive index is modified using electro-optic effect, where the signal slows down and a $\pi$ phase shift is obtained in case '1' is applied. Hence, depending on the applied voltage, constructive and destructive interference can be obtained when both signals are combined at the output. This effect has been demonstrated to modulate signals at 40 Gbp/s, under 4.5dB insertion loss IL (the fraction of optical power lost) and 3.2dB extinction ratio ER [10], i.e., the ratio of the output power when logic level '1' is transmitted (ON state) to the output power when logic level '0' is transmitted (OFF state).

*MRR as Modulator:* Fig. 2(b) illustrates a modulator implemented using an MRR controlled by a voltage applied to its Positive-Intrinsic-Negative (PIN) junction [11]. In the initial state (i.e., no voltage is applied), the MRR resonant wavelength is set to $\lambda_0$. This leads to the coupling of the light at wavelength $\lambda_0$ into the ring, which results in a small fraction of signal power transmitted. When a voltage is applied, the refractive index of the MRR is blue shifted, i.e., most of the input signal power is transmitted to the output. Equation (2) is the through transmission $\varphi_t$ of the MRR modulator to the output as defined in [12].

$$\varphi_t(\lambda_{signal}, \lambda_{res}) = \frac{a^2(\lambda_{res})r_2^2 - 2a(\lambda_{res})r_1r_2 \cos[\theta(\lambda_{signal}, \lambda_{res})] + r_1^2}{1 - 2a(\lambda_{res})r_1r_2 \cos[\theta(\lambda_{signal}, \lambda_{res})] + [a(\lambda_{res})r_1r_2]^2} \quad (2)$$

where $r_1$ and $r_2$ are the self-coupling coefficients, $\lambda_{res}$ and $\lambda_{signal}$ are the MRR resonant wavelength and signal wavelength, respectively. $\Delta\lambda$ is the wavelength shift between ON and OFF states, $a$ is the single-pass amplitude transmission, and $\theta$ is the single-pass phase shift.

*MRR as All-Optical Filter:* Fig. 2(c) illustrates an optically controlled MRR. Using two-photon absorption (TPA) effect, a shift of the ring refractive index is obtained by applying a high intensity pump signal at $\lambda_{pump}$ [13]. The wavelength of the pump signal is slightly detuned from the filter resonant wavelength (0.1nm in [14]). In order to avoid crosstalk with the modulated signal, the next resonance $\lambda_{ref}$ of the MRR is used for the filtering operation ($\lambda_{ref}=\lambda_{pump}+FSR$). The resonant wavelength is blue shifted, as illustrated in the figure. In case no pump signal is applied, signals at $\lambda_0$ and $\lambda_1$ continue propagating on the waveguide. When a power pump signal is injected, the resonant wavelength of the filter is shifted to $\lambda_1$, which leads a transmission of the corresponding signal to the drop port. The drop transmission $\varphi_d$ is given in Eq. (3).

$$\varphi_d(\lambda_{signal}, \lambda_{res}) = \frac{a(\lambda_{res})(1 - r_1^2)(1 - r_2^2)}{1 - 2a(\lambda_{res})r_1r_2 \cos[\theta(\lambda_{signal}, \lambda_{res})] + [a(\lambda_{res})r_1r_2]^2} \quad (3)$$

According to [13], the effective index $n_{eff}$ is

$$n_{eff} = n_0 + n_2 P/S \quad (4)$$

where $n_0$ and $n_2$ are the linear and non-linear refractive indexes, respectively. $P$ is the pump signal power and $S$ is the effective cross-sectional area of the filter. In [14], a 0.1nm shift was reported for an average 10mW pump signal. Using this non-linear optical effect, a fully optical AND gate with 100ps switching time was fabricated [15]. In this paper, we rely on this physical effect to design an optical multiplexer.

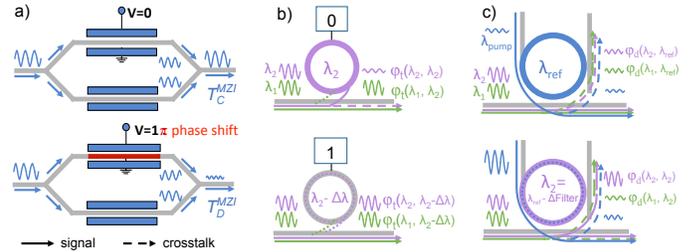

**Fig. 2:** Optical devices. a) MZI in constructive and destructive states, b) MRR in ON and OFF states, and c) all-optical add-drop filter.

These devices have been investigated in the design of integrated optical computing circuits. MZI, the most mature device among the three listed above, allows implementing microwave filters processors [5]. Whereas, MRRs were used to design optical lookup table (OLUT) [6] and a reconfigurable directed logic architecture (RDL) [16]. In [17], an all-optical reconfigurable circuit is implemented to perform logic operations using MRRs. The circuit relies on a multiplexer that selects the logic function by applying the required pump signal.

Differently from prior works, we address the design of a stochastic circuit using integrated optics, which has never been investigated so far.

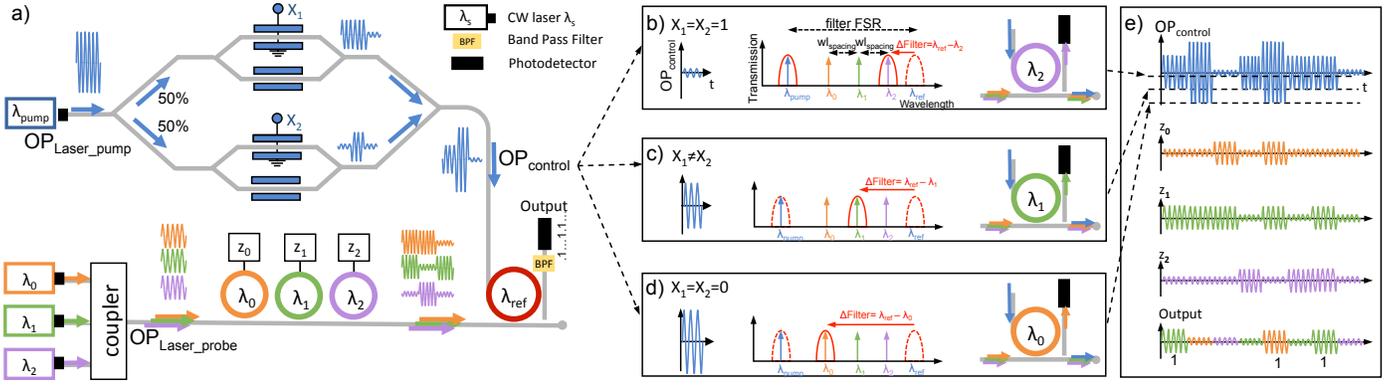

Fig. 3: Optical SC unit of a 2$^{nd}$ order polynomial function. a) The optical circuit. The transmissions of the signals at $\lambda_2$, $\lambda_1$, and $\lambda_0$ to the drop port of the filter are shown in b), c), and d), respectively. e) Transmission example of the probe and control signals, and power received by the photodetector.

## III. PROPOSED DESIGN

In this section, we first illustrate the proposed optical SC architecture with an example. The general architecture is then presented and key design challenges are detailed.

### A. Architecture Overview

The optical circuit is composed of an adder and a multiplexer, similarly to the ReSC circuit introduced in Section II. The adder contains MZIs controlled by data $x_i$. A high power optical signal input ($OP_{Laser\_pump}$) is emitted by a continuous wave (CW) laser source, where its power is equally distributed to MZI devices. The output is an optical control ($OP_{control}$) signal for which insensitivity is modulated according to the combination of destructive and constructive interference occurring in the MZIs. The choice of using MZI in this part of the design is based on the non-resonant structure characteristic of the MZI. In other words, it is not affected by high power signals compared to MRR

The multiplexer is implemented using an all-optical add-drop filter receiving optical signals modulated by coefficients $z_j$. The resonant wavelength of the filter depends on the intensity of the pump signal output by the adder. By controlling the resonant wavelength of the filter, it is possible to extract a coefficient signal, thus implementing the multiplexing operation. The output signal is transmitted to a Band Pass Filter (BPF) for pump signal absorption and is eventually received by a photodetector, where E/O conversion is carried out.

Fig. 3(a) illustrates a 2$^{nd}$ order Bernstein polynomial function. Three MRRs modulate optical signals at wavelengths $\lambda_0$, $\lambda_1$, and $\lambda_2$ according to coefficients $z_0$, $z_1$, and $z_2$, respectively. The filter is initially tuned to the resonant wavelength $\lambda_{ref}$. The adder is composed of two MZIs controlled by data $x_1$ and $x_2$. According to the possible combination of $x_1$ and $x_2$, three optical power levels of the pump signal can be obtained, which lead to the following scenarios:

- $x_1=x_2=1$ (Fig. 3(b)): Both MZIs operate in the destructive state. Therefore, the optical power is strongly attenuated and the filter is tuned to $\lambda_2$ (the right most resonant wavelength). As a result, the coefficient signal at wavelength $\lambda_2$ is dropped to the output.

- $x_1 \neq x_2$ (Fig. 3(c)): One of the MZIs operates in the constructive state. This results in a higher transmission of the laser signal to the filter (approximately half of the power). The filter resonant wavelength is shifted to $\lambda_1$, thus leading to the transmission of signals at $\lambda_1$ to the output.

- $x_1=x_2=0$ (Fig. 3(d)): Both MZIs operate in the constructive state, which leads to the maximum power transmission to the filter. The filter resonant wavelength is tuned to the transmission of the left most signal wavelength, i.e., $\lambda_0$.

Hence, while the filter resonant wavelength is initially set to $\lambda_{ref}$, a non-linear effect achieved through the pump excitation leads to drift in the resonance wavelength to $\lambda_0$, $\lambda_1$ and $\lambda_2$. The photodetector thus receives coefficient $z_0$, $z_1$ and $z_2$, according to the value of data $x_1$ and $x_2$, as illustrated in Fig. 3(e). Eventually, the number of ones is counted at the receiver side to complete the operations needed for stochastic computing.

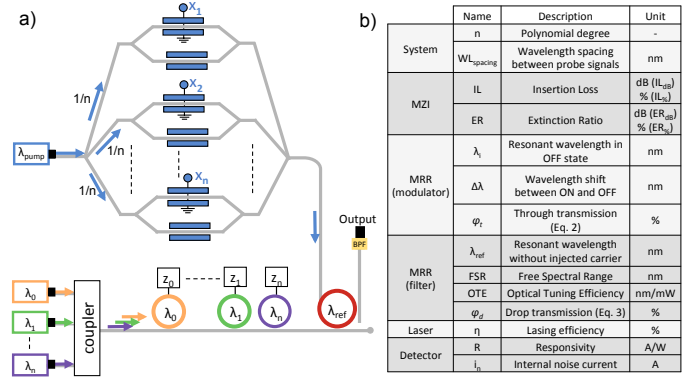

Fig. 4: a) Generic architecture for optical stochastic computing circuit, and b) System-level and device-level parameters.

### B. Generic Circuit and Design Challanges

The architecture we propose is generic and can be implemented for an $n$-order Bernstein polynomial function, as illustrated in Fig. 4(a). It involves n MZIs and $n+1$ MRRs to modulate the data and the coefficients, respectively. The optical power of the pump laser is equally distributed to the MZIs using $n$-outputs and $n$-inputs splitter and combiner, respectively. The use of Wavelength Division Multiplexing (WDM) for the coefficient signals propagation involves $n+1$ probe lasers at wavelengths $\lambda_i$ ($0 \leq i \leq n$) separated by wavelength spacing $WL_{spacing}$ (i.e., the wavelength difference between consecutive signals).

The design of such architecture is challenging due to the heterogeneity of the involved devices and hence the physical effects. Its optimization requires considering both technological and system-level parameters (Fig. 4(b)) and leads to conflicting objectives. For instance, low $WL_{spacing}$ is mandatory to reduce the optical power required to control the filter. However, a low spacing also increases the crosstalk between coefficient signals, which in turn requires increasing the probe signals power. There is a tradeoff to explore between the pump laser and the probe lasers. The optimization of the architecture thus calls for a comprehensive transmission model, which we formalize in Section IV.

## IV. IMPLEMENTATION AND MODEL

To analyze the proposed design, an analytical model is developed. It allows evaluating key design metrics such as the Signal-to-Noise Ratio (SNR) and laser power, taking into account both technological and system-level parameters.

### A. Transmission Model

We first define $WL_{spacing}$ as the wavelength distance between two consecutive probe signals:

$$WL_{spacing} = \lambda_{i+1} - \lambda_i \quad (5)$$

The transmission of a probe signal at $\lambda_i$ (which corresponds to coefficient $z_i$ in the ReSC architecture) is presented in Eq. (6). As already defined in Section II, $\varphi_t$ and $\varphi_d$ correspond to MRR through and filter drop transmissions respectively. The transmission of '1' for coefficient $z_i$ leads to $\Delta\lambda$ detuning of the modulating $MRR_i$ (ON state). The signal also propagates through the other MRRs, experiencing attenuation depending on their modulation states (which depends on the values of the remaining coefficients $z_w$). The signal is then dropped to the output by the filter, experiencing a transmission depending on the detuning $\Delta Filter$. In case '0' is transmitted ($z_i=0$), the modulating MRR is tuned to the signal wavelength (OFF-state).

$$T_{s,z}[i] = \underbrace{\varphi_t(\lambda_i, \lambda_i - \Delta\lambda \times z_i)}_{\text{Transmission through the modulating MRR}} \times \underbrace{\prod_{\substack{w=0\\w \neq i}}^{n} \varphi_t(\lambda_i, \lambda_w - \Delta\lambda \times z_w)}_{\text{Transmission through the other modulators}} \times \underbrace{\varphi_d(\lambda_i, \lambda_{ref} - \Delta Filter(x))}_{\text{Transmission through the filter}} \quad (6)$$

The transmission of the probe signals to the output depends on the filter for which the initial resonant wavelength is $\lambda_{ref}$ (i.e., the resonant wavelength in case no control power is applied). The filter detuning is defined as:

$$\Delta Filter(x) = OP_{Laser\_pump} \times OTE \times \frac{1}{n}\sum_{i=1}^{n} T_i^{MZI}[x_i] \quad (7.a)$$

$$\begin{cases} T_i^{MZI}[x_i] = IL_\% & x_i = 0 \\ T_i^{MZI}[x_i] = IL_\% \times ER_\% & x_i = 1 \end{cases} \quad (7.b)$$

where $OTE$ is the optical tuning efficiency (expressed in nm/mW). The detuning depends on the total transmission of $OP_{Laser\_pump}$ through the $n$ parallel MZIs. The transmission through each MZI depends on the corresponding modulated data $x_i$, where '0' and '1' lead to constructive and destructive states of the MZI, respectively Eq. (7.b). $IL_{dB}$ and $ER_{dB}$ are the conversion results of the ratio to dB of $IL_\%$ and $ER_\%$ respectively. The pump signal absorption induced by the BPF is neglected in our model. SNR is eventually estimated as follows:

$$SNR = OP_{Laser\_probe} \times \frac{R}{i_n} \times \left( T_{s,z_i=1}[i] - \sum_{\substack{w=0\\w \neq i}}^{n} T_{s,z_w=1}[w] \right) \quad (8)$$

where $i_n$ and $R$ are the photodetector internal noise and responsivity, respectively. $T_{s,z_i=1}[i]$ is the transmission of signal $i$ as '1' while the remaining signals are '0'. $T_{s,z_w=1}[w]$ is the transmission of the crosstalk signal $w$ as '1' for the same signal $i$ when transmitted as '0'. Eq. (9) gives the Bit-Error-Rate (BER) assuming On/Off key modulation (OOK) of the probe signals.

$$BER = \frac{1}{2} erfc\left(\frac{SNR}{2\sqrt{2}}\right) \quad (9)$$

### B. Design Methods

The performance and the energy efficiency of the architecture depend on the numerous devices characteristics and on the combination of the related parameters. Due to the large design space to explore, methods are needed to guide designers in the optimization process. In the following, the methods **MRR-first** and **MZI-first** are briefly introduced.

**MRR-first:** allows exploring MZI characteristics and minimizes the required pump laser power $OP_{Laser-pump}$ according to MRRs parameters. For this purpose, the MRR resonant wavelengths $\lambda_i$ are first defined according to $WL_{spacing}$. The transmission $T_{s,z}[i]$ then allows estimating the worst-case SNR for a given probe laser power $OP_{Laser\_probe}$, or to find the minimum laser power needed to reach a given SNR. Then, according to a filter resonant wavelength $\lambda_{ref}$ and a MZI insertion loss IL, the minimum pump power is computed. Eventually, the extinction ratio ER is given by the pump signal attenuation required to tune the filter to $\lambda_n$, the right-most signal wavelength.

**MZI-first**: allows exploring MRRs characteristics and to minimize the required probe laser power. For this purpose, the pump laser power and the MZI IL and ER are defined. This allows estimating the power level of the control signal to tune the filter: for a given $\lambda_{ref}$, it is possible to define $\lambda_i$ and vice-versa. Eventually, SNR and laser probe power can be defined according to the objective (power, robustness, speed).

## V. RESULTS

In this section, we first detail the design of the 2$^{nd}$ order circuit illustrated in Fig. 3. Then, the influence of MZI characteristics on the laser probe power requirements is evaluated. We then investigate the use of pulse-based lasers to maximize the energy efficiency of the circuit. Finally, we discuss several extension opportunities of this work.

### A. Design of 2$^{nd}$ Order Optical Stochastic Circuit

In the first experiments, we design a 2$^{nd}$ order polynomial circuit using the **MRR-first** method. For this purpose, we define $WL_{spacing}$=1nm and $\lambda_2$=1550. Fig. 5(a) and (b) illustrate the transmission of the MRRs and the filter as well as the probe signals (represented with vertical arrows).

In Fig. 5(a), we assume $z_0$=0, $z_1$=1, and $z_2$=0. $MRR_1$ is thus slightly detuned, which leads to a higher transmission of the signal at $\lambda_1$, whereas $\lambda_0$ and $\lambda_2$ are attenuated. We assume $x_1$=$x_2$=1, which leads to a resonant wavelength of the filter at $\lambda_2$. As a result, the total transmission (i.e., the transmission through all the MRRs and the filter) of the signals at $\lambda_2$, $\lambda_1$ and $\lambda_0$ are 0.091, 0.004 and 0.0002 respectively. The photodetector receives an optical power corresponding to the sum of the power of the three signals. By assuming 1mW for $OP_{Laser\_probe}$, a total power of 0.0952mW is received. In Fig. 5(b), we assume $z_0$=1, $z_1$=1, $z_2$=0 and $x_1$=$x_2$=0, which leads to a detuning of $MRR_0$ and $MRR_1$, an attenuation of $\lambda_2$, and the tuning of the filter resonant wavelength at $\lambda_0$. Therefore, the total transmission of the signal at $\lambda_0$ is 0.476 and the power received by the detector is 0.482mW.

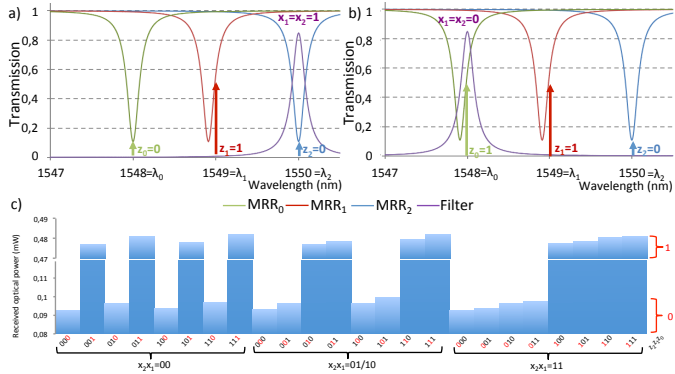

Fig. 5: Transmission of MRRs and filter. a) probe signal at $\lambda_2$ is transmitted as 0, b) probe signal at $\lambda_0$ is transmitted as 1, and c) The optical power received by the photodetector for all input combinations.

The de-randomizer implies to identify data '1' from data '0'. In the optical domain, it involves to associate power levels to the transmitted data value. We thus estimate the power received at the photodetector for all combinations of data ($x_1$ and $x_2$) and coefficients ($z_0$, $z_1$, and $z_2$) and for a 1mW laser probe. As reported in Fig. 5-c, data '0' and '1' lead to received optical power in the ranges of 0.092-0.099mW and 0.477-0.482mW, respectively. This allows a correct execution of SC in the optical domain, thus validating the proposed circuit.

We estimate the pump laser power still following the **MRR-first** method. For this purpose, we assume $\lambda_{ref}$=1550.1nm, which corresponds to a 0.1nm shift compared to $\lambda_2$ and which contributes to reduce the pump power due to the shorter wavelength ranges. We assume 0.1nm/10mW for OTE [14]. By considering $IL_{dB}$=4.5dB [10], we estimate that the minimum pump power required to reach $\lambda_0$ (i.e., case $x_1$=$x_2$=0, Fig. 5(b)) is 591.8mW. $ER_{dB}$ of 13.22dB is obtained by evaluating the required pump signal power attenuation to detune the filter to $\lambda_1$ and $\lambda_2$.

### B. Influence of MZI Characteristics on Probe Laser Power

Following the **MZI-first** method, minimum laser probe powers are evaluated by considering ranges of values for ER and IL typically observed in the literature [18] and [19]. In this study, we assume a $2^{nd}$ order polynomial function.

Fig. 6 illustrates the results for $OP_{Laser\_pump}$=0.6W and BER=$10^{-6}$. By assuming the MZI device in [19] ($IL_{dB}$=6.5 and $ER_{dB}$=7.5), the required laser probe power would be 0.26mW. Obviously, the minimum value of $OP_{Laser\_probe}$ rises with the increase in $IL_{dB}$ and the reduction of $ER_{dB}$, which is explained as follows: the lower the total transmission in the MZIs, the smaller the wavelength spacing and the higher the signal crosstalk. Increasing the probe laser power not only has a negative effect on the circuit energy efficiency, but it can also induce non-linear effect in the filter, which would lead to undesired shift of its resonant wavelength. This could be avoided by increasing the pump power instead, which leads to a design trade-off involving the power of the pump and probe signals. We also evaluate the opportunities for laser power reduction by leveraging constraints of the optical signal transmission robustness.

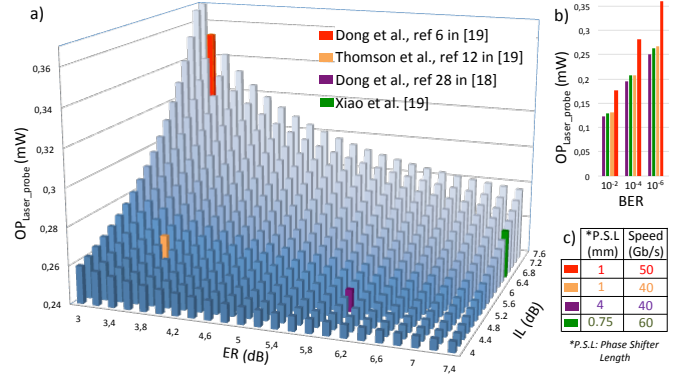

Fig. 6: Minimum probe laser power according to a) $IL_{dB}$ and $ER_{dB}$ for $10^{-6}$ BER b) targeted BER, and c) MZIs speed and phase shifter length.

As illustrated in Fig. 6(b), targeting $10^{-2}$ BER instead of $10^{-6}$ leads to a 50% power reduction. The lack of accuracy in the optical domain could be alleviated by transmitting longer streams of bits in the stochastic domain. This also allows exploring a tradeoff between computing accuracy and transmission robustness, which involve device characteristics related to the speed and the area (Fig. 6(c)). For instance, a high modulation speed (e.g., 60Gb/s [19]) and a high laser power could be combined to reduce the bit-streams transmission rate, thus maximizing the circuit throughput.

### C. Energy Breakdown using Pulse-Based Pump Laser

In order to maximize the energy efficiency of the design, we consider the use of a laser generating 26ps pump pulses [15]. The energy efficiency for orders 2, 4 and 6 are investigated under the following assumptions: 1Gb/s modulation speed for MZI and MRRs, 20% lasing efficiency and [0.1nm-0.3nm] wavelength spacing range.

Fig. 7(a) gives the energy consumption per laser type (i.e. one curve for pump laser and another curve for $n$ probe lasers) according to the wavelength spacing. As seen from the figure, the energy consumed by the pump laser and the probe lasers follow opposite trends: i) for $WL_{spacing}$ smaller than 0.165nm, the total energy consumption is dominated by the probe lasers (which is required to compensate crosstalk effects); ii) for $WL_{spacing}$ larger than 0.165nm, the pump laser power dominates, which is due to the larger wavelength shift occurring in the filter. There is thus an optimal wavelength spacing value minimizing the total energy consumed by the lasers. Interestingly, the optimal wavelength spacing is independent from the polynomial degree, which we could take advantage of to design a reconfigurable circuit that executes polynomial functions of various orders.

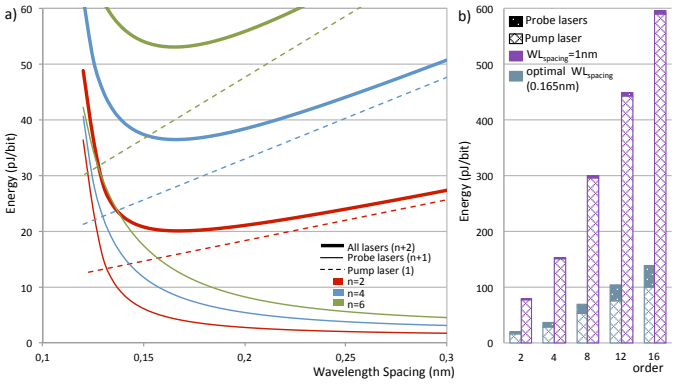

**Fig. 7:** Laser energy consumption per computed bit according to a) WL$_{spacing}$ and b) the polynomial degree.

In order to evaluate the scalability of the architecture, we evaluate the total laser energy consumption according to *n*, assuming 1nm and optimal wavelength spacing. As illustrated in Fig. 7(b), the use of optimal WL$_{spacing}$ leads to 76.6% energy saving. From these results, it is possible to estimate a circuit power consumption and throughput, taking into account the required polynomial degree. For instance, Gamma correction application, which is a non-linear function used in image processing, involves a 6$^{th}$ order degree. Compared to the 100MHz frequency considered in [9], the use of integrated optics will lead to a 10x speedup. It is also worth mentioning that power density limitation could be leveraged using a parallel implementation of the architecture.

### D. Discussion and Future Work

While this work is the very first addressing the optical implementation of stochastic computing circuits, it leads to new opportunities but also to new challenges. As it was shown, the use of pulse-based lasers leads to significant energy reduction; however, it also requires synchronization on the detector side to read the received signals only during the short light emission. This calls for feedback loop-based control circuit involving monitoring and voltage/thermal tuning for device calibration. The design of such circuit relies on energy-area tradeoff we plan to explore.

We have also highlighted that the optical transmission robustness leads to throughput-accuracy tradeoff, which we believe is particularly interesting for SC applications. Indeed, this would allow adapting the length of the bit-streams and its transmission speed according to application real-time constraints and energy requirements. In order to carry out this study, we plan to develop a SPICE model for transient simulation of the optical circuit.

Another direction of future work is the implementation of the SC interfaces (namely randomizer and de-randomizer) in the optical domain which will allow comparing our optical SC circuit with other CMOS-based SC circuits. For this purpose, we will investigate the use of compact chaotic lasers [20] for ultra-fast generation of random bits. Regarding the probability computation of the output bit-streams, the benefits of using high responsivity avalanche photodiode [21] will be evaluated.

## VI. CONCLUSION

This paper presents the first work addressing the implementation of Stochastic Computing (SC) in the optical domain. We show that a circuit implementing a 2$^{nd}$ order polynomial degree function and operating at 1Ghz leads to 20.1pJ laser consumption per computed bit. As a key result, we also show that the wavelength spacing leading to optimal laser power consumption is independent from the polynomial degree. This will be taken advantage of for the design of a reconfigurable version of our architecture. As detailed in the paper, future works are numerous and involve i) the design of a controller for circuit monitoring, synchronization and calibration, ii) the exploration of throughput-accuracy tradeoff using transient simulations and iii) the implementation in the optical domain of randomizer and de-randomizer blocks.